%% file: SPAWC25_DE_SB_IDMA.tex
\documentclass[twocolumn,conference,10pt,letterpaper,final]{IEEEtran}
\usepackage{amsthm,amsmath,amsfonts,amsbsy}
\usepackage{xcolor}
\usepackage{mathrsfs}
\usepackage{mathtools}
\usepackage{bm}
\usepackage{balance}
\usepackage{verbatim}
\usepackage{tikz}
\usetikzlibrary{patterns}
\usepackage{pgfplots}
\usepackage{scalerel}
\usepackage{algorithm}
\usepackage{lipsum}
\usepackage{cite}
\usepackage[nolist]{acronym}
\newcommand{\real}{\mathbb{R}}
\newcommand{\PUPE}{\mathsf{PUPE}}
\newcommand{\inalpha}{\mathcal{X}}

\newcommand{\code}{\mathcal{C}}
\newcommand{\pset}{\mathcal{P}}
\newcommand{\aset}{\mathcal{A}}
\newcommand{\npre}{n_{\scaleto{\mathsf{PRE}}{3.5pt}}}
\newcommand{\ypre}{\bm{y}_{\scaleto{\mathsf{PRE}}{3.5pt}}}
\newcommand{\npo}{n_{\scaleto{0}{3.5pt}}}
\newcommand{\expect}[1]{\mathsf{E}\!\left[#1\right]}
\newcommand{\prob}[1]{\mathsf{P}\!\left[#1\right]}
\newcommand{\poisson}[1]{\mathrm{Pois}\!\left(#1\right)}
\newcommand{\graph}{\mathcal{G}} 
\newcommand{\un}{\mathtt{u}} 
\newcommand{\sn}{\mathtt{s}}
\newcommand{\mzero}[2]{m_{\scaleto{#1}{5pt}}^{\scaleto{(#2)}{6pt}}}
\newcommand{\msg}[3]{m_{\scaleto{#1 \rightarrow #2}{5pt}}^{\scaleto{(#3)}{6pt}}}
\newcommand{\msgdum}[2]{m_{\scaleto{#1 \rightarrow #2}{4pt}}}
\newcommand{\neigh}[1]{\mathscr{N}\!\left(#1\right)}
\newcommand{\seg}[2]{\bm{#1}{\scaleto{[#2]}{7pt}}}
\newcommand{\segx}[2]{\bm{#1}^{(i)}{\scaleto{[#2]}{7pt}}}
\newcommand{\dec}[1]{\mathsf{dec}\!\left[#1\right]}
\newcommand{\flag}{\scaleto{\mathtt{ERR}}{5pt}}
\newcommand{\missing}{\scaleto{\mathtt{MISSING}}{5pt}}
\newcommand{\decoded}{\scaleto{\mathtt{DECODED}}{5pt}}
\newcommand{\state}{\scaleto{\mathtt{STATE}}{5pt}}
\newcommand{\hash}{\mathsf{h}}

\newcommand{\vnode}[2]{\node[circle,fill=black,draw=black,thin,minimum height = 0.2cm, minimum width = 0.2cm, inner sep = 0] (#2) at (#1) {};}
\newcommand{\cnode}[2]{\node[rectangle,fill=black,draw=black,thin,minimum height = 0.2cm, minimum width = 0.2cm, inner sep = 0] (#2) at (#1) {};}

\begin{acronym}
	\acro{2SRA}{two-step random access}
	\acro{4SRA}{four-step random access}
	\acro{5GNR}{5G New Radio}
	\acro{AWGN}{additive white Gaussian noise}
	\acro{APP}{a posteriori probability}
	\acro{BEC}{binary erasure channel}
	\acro{BP}{belief propagation}
	\acro{BS}{base station}
	\acro{CN}{check node}
	\acro{DE}{density evolution}
	\acro{DT}{dependency testing}
	\acro{i.i.d.}{independent, identically-distributed}
    \acro{IDMA}{interleaver division multiple access}
	\acro{IoT}{Internet of Things}
	\acro{LDPC}{low-density parity-check}
	\acro{LLR}{log-likelihood ratio}
	\acro{MAC}{multiple access}
	\acro{MAP}{maximum a posteriori probability}
	\acro{ML}{maximum likelihood}
	\acro{mMTC}{massive machine-type communication}
    \acro{MUD}{multiuser detection}
	\acro{OFDM}{orthogonal frequency-division modulation}
	\acro{OMP}{orthogonal matching pursuit}
	\acro{PRACH}{physical random access channel}
	\acro{PUPE}{per-user probability of error}
	\acro{r.v.}{random variable}
	\acro{SB-IDMA}{sparse block interleaver division multiple access}
	\acro{SCL}{successive cancellation list}
	\acro{SIC}{successive interference cancellation}
	\acro{SN}{slot node}
	\acro{SNR}{signal-to-noise ratio}
	\acro{UMAC}{unsourced multiple access}
	\acro{TIN}{treat-interference-as-noise}
	\acro{UN}{user node}
	\acro{UT}{user terminal}
	\acro{VN}{variable node}
\end{acronym}

\IEEEoverridecommandlockouts

\begin{document}
\title{Density Evolution Analysis of Sparse-Block IDMA}
\author{
	\IEEEauthorblockN{Jean-Francois Chamberland}
	\IEEEauthorblockA{\textit{Texas A\&M}\\
		chmbrlnd@tamu.edu}
    \and
	\IEEEauthorblockN{Gianluigi Liva}
	\IEEEauthorblockA{
		\textit{German Aerospace Center (DLR)}\\
		gianluigi.liva@dlr.de}\thanks{G.L. acknowledges the financial support by the Federal Ministry of Research, Technology and Space of Germany in the programme of ``Souver\"an. Digital. Vernetzt.'' Joint
		project 6G-RIC, project identification number: 16KISK022.}
        \and
        \IEEEauthorblockN{Krishna Narayanan}
		\IEEEauthorblockA{
			\textit{Texas A\&M}\\
			krn@ece.tamu.edu}
}

\maketitle

\setlength{\abovedisplayskip}{7.7pt}
\setlength{\belowdisplayskip}{7.7pt}
\setlength{\abovedisplayshortskip}{5.7pt}
\setlength{\belowdisplayshortskip}{5.7pt}

\begin{abstract}
	Sparse block interleaver division multiple access (SB-IDMA) is a recently introduced unsourced multiple access protocol that aims to improve the performance of the grant-free two-step random access transmission protocol of the 3GPP 5G New Radio standard. We introduced a density evolution analysis of the successive interference cancellation receiver of SB-IDMA, providing a theoretical characterization of its performance.  
\end{abstract}


\section{Introduction}\label{sec:intro}
The \ac{2SRA} protocol was recently introduced in Release 16 of the 3GPP \ac{5GNR} standard \cite{5GNR16} to reduce the latency of the random access channel. Despite implementing a grant-free access mechanism, \ac{2SRA} falls short of providing the energy-and-spectrum efficiency required by \ac{mMTC} and \ac{IoT} networks \cite{LP24}. In \cite{whitepaper24}, \ac{SB-IDMA} was introduced as an evolution of \ac{2SRA}, aiming at improving its energy efficiency and at enabling the support of large \ac{UT} populations. \ac{SB-IDMA} builds on the \ac{UMAC} paradigm introduced in \cite{Pol17}, implementing a two-phase approach where, as for \ac{2SRA}, the active \acp{UT} transmit first a preamble, and then data units in time/frequency resources pointed by the selected preamble. \ac{2SRA} adopts, for the second phase, a slotted Aloha \cite{Rob75} approach. \ac{SB-IDMA} employs in the second phase a more sophisticated transmission strategy, which is inspired by the sparse \ac{IDMA} scheme of \cite{PAV+22}. Under a straightforward \ac{TIN} \ac{SIC} receiver architecture, \ac{SB-IDMA} can achieve a performance that approaches the finite-length achievability bound of \cite{Pol17} over the Gaussian \ac{UMAC} channel up to moderately large system loads \cite{whitepaper24}.  In this paper, we study the performance of \ac{SB-IDMA} under \ac{TIN}-\ac{SIC} decoding by means of \ac{DE} analysis \cite{RU08}. The analysis addresses the second phase of the \ac{SB-IDMA} transmission. It relies on a graphical model of the iterative \ac{SIC} process, considering the asymptotic limit of large frame sizes while retaining the finite size of the user payload transmissions. The analysis provides useful insights into the behavior of \ac{SB-IDMA} under \ac{TIN}-\ac{SIC} decoding.

\section{Preliminaries}\label{sec:preliminaries}

We use capital letters for random variables, e.g., $X$, and lowercase letters for their realizations, e.g., $x$. We denote by $\poisson{\lambda}$ the $\lambda$-mean Poisson distribution, and by $\mathcal{N}(\mu,\sigma^2)$ the Gaussian distribution with mean $\mu$ and variance $\sigma^2$. Finally, we use the shorthand $[x]^+=\max(0,x)$.

We consider the  Gaussian \ac{UMAC} model of \cite{Pol17}. We denote by $\inalpha \subseteq \real$ the alphabet used for transmission, i.e., the set of symbols that can be used by the transmitter. We assume that the user population comprises $K$ \acp{UT}, out of which $K_a\ll K$ are active. Each active user attempts the transmission of $k$ information bits. We consider transmission over $n$ channel uses, and we refer to $n$ as the \ac{UMAC} frame length. 
The channel output (for a single channel use) is
\[
Y = \sum_{i = 1}^{K_a} X^{(i)} + Z
\]
where $X^{(i)} \in \inalpha$ is the symbol transmitted by the $i$th active \ac{UT}, and $Z$ is the \ac{AWGN} term, $Z \sim \mathcal{N}(0,\sigma^2)$. 
The sequence of symbols transmitted by the $i$th user is $\bm{X}^{(i)} = ( X^{(i)}_1, X^{(i)}_2, \ldots, X^{(i)}_n)$.
We enforce the power constraint  $||\bm{X}^{(i)}||_2^2\leq nP$.
The per-user \ac{SNR} is
$E_b/N_0 = {nP}/{2k\sigma^2}$
where $E_b$ is the energy per information bit, and $N_0$ is the single-sided noise power spectral density.
All users adopt the same \ac{UMAC} code. In the idealized case where the receiver knows the number of active users, the decoder outputs a list $\mathcal{L}$ of $K_a$ codewords. The \ac{PUPE} is
\begin{equation}
	\PUPE := \frac{1}{K_a}\sum_{i = 1}^{K_a} \prob{\bm{X}^{(i)} \notin \mathcal{L}}.
	\label{eq:PUPE}
\end{equation}
We finally denote the user density as $\mu := K_a/n$.

\section{Sparse-Block IDMA}\label{sec:SB-IDMA}

The reference frame structure of \ac{SB-IDMA} is depicted in Fig.~\ref{fig:SBIDMA} \cite{whitepaper24}.
Following the \ac{2SRA} protocol introduced in \cite{5GNR16}, we divide the \ac{UMAC} frame in two parts: a preamble transmission part (PRACH - \emph{physical random access channel}, in the 3GPP terminology), composed by $\npre$ channel uses, and a data transmission part, of length $n-\npre$ channel uses. The data part is further partitioned in $N$ slots (\emph{physical uplink shared channel occasions}, PO). Each slot comprises $\npo$ channel uses. It follows that the frame length is $
n = \npre + N\times \npo$. 

\subsection{Transmission}\label{sec:SB-IDMA:TX}
Each \ac{UT} is equipped with a preamble dictionary $\pset$, a set $\aset$ of access patterns, a hash function $\hash$, and a block code $\code$. The sets $\pset$, $\aset$, the hash function, and the block code are the same for all \acp{UT}. The preamble length is $\npre$. There is a one-to-one correspondence between preambles and access patterns, hence the number of preambles $|\pset|$ equals the number of access patterns $|\aset|$. An access pattern $\bm{a} = (a_1, a_2, \ldots, a_{N_s})$ is a sequence of $N_s$ distinct indexes in $\{1,2,\ldots,N\}$. The hash function $\hash$ takes as input a bit sequence, and it outputs a preamble index in  $\{1,2,\ldots,|\pset|\}$. The block code $\code$ is composed of $M=2^k$ codewords of length $n_c$.

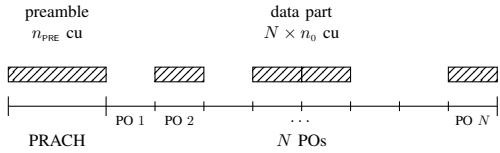
\begin{figure}
	\centering
	\input{diagramm.tikz}
	\vspace{-2mm}
	\caption{\ac{SB-IDMA} frame structure.}
	\vspace{-2mm}
	\label{fig:SBIDMA}
\end{figure}

The transmission at each \ac{UT} works as follows (see Fig.~\ref{fig:SBIDMA}).

\smallskip

\noindent\textbf{Step 1.} The $k$-bits message $\bm{u}$ is hashed, generating an index $\hash(\bm{u})$ which uniquely identifies (a) a preamble $\bm{p}$ in $\pset$,  and (b) the corresponding access pattern $\bm{a}$ in $\aset$.

\smallskip

\noindent\textbf{Step 2.} The message $\bm{u}$ is encoded via the block code $\code$, resulting in the codeword $\bm{c} = (c_1, c_2, \ldots, c_{n_c})$.

\smallskip

\noindent\textbf{Step 3.} The codeword is split into $N_s$ segments of length $\npo$, $\bm{c} = (\seg{c}{1},\seg{c}{2}, \ldots, \seg{c}{N_s})$. 

\smallskip

\noindent\textbf{Step 4.} The preamble $\bm{p}$ is transmitted over the PRACH, and $\seg{c}{j}$ ($j=1,2,\ldots,N_s$) is transmitted in the PO pointed by $a_j$.

\subsection{Detection and Decoding}\label{sec:SB-IDMA:RX}
At the \ac{BS}, the receiver observes the noisy superposition of the $K_a$ active user transmissions
\begin{equation}
	\bm{y} = \sum_{i=1}^{K_a} \bm{x}^{(i)} + \bm{z}.
\end{equation}
We partition  $\bm{y} = (\ypre,\seg{y}{1}, \seg{y}{2}, \ldots, \seg{y}{N})$
 where $\ypre$ is the sequence of $\npre$ channel outputs associated with the PRACH, and $\seg{y}{j}$ is the sequence of $\npo$ channel outputs associated with the $j$th PO. The \ac{TIN}-\ac{SIC} receiver operates as follows.

\smallskip

\noindent\textbf{Step 1.} A preamble detector outputs a list $\mathcal{D}$ of preambles.

\smallskip

\noindent\textbf{Step 2.} The following procedure is performed for all the preambles detected in Step 1: Consider a preamble in $\mathcal{D}$, and denote its corresponding access pattern $\bm{a}$. A sequence of $n_c$ observations is obtained as  $\seg{y}{\bm{a}}:=(\seg{y}{a_1},\seg{y}{a_2},\ldots,\seg{y}{a_{N_s}})$
	and it is input to the decoder of $\code$. If decoding succeeds\footnote{Here, we assume that the decoder possesses some error detection capability, i.e., the decoder can output an error flag. See \cite{whitepaper24} for details.}, the interference contributions of the decoded codeword and of the corresponding preamble are removed from $\bm{y}$.

\smallskip

Steps $1$-$2$ are repeated for a maximum number of iterations.

\subsection{Performance}\label{sec:SB-IDMA:PUPE}
\begin{figure}[!t]
	\centering
	\includegraphics[width=0.8\columnwidth]{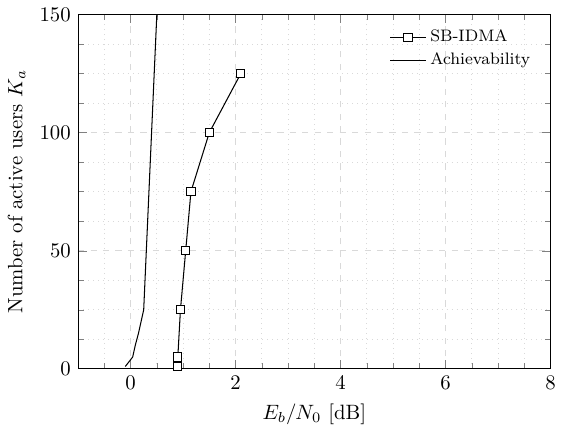}
    \vspace{-2.5mm}
	\caption{Number of active users vs. \ac{SNR} required to achieve a $\PUPE=5\times 10^{-2}$. The frame length is $30000$ channel uses.}
	\vspace{-2mm}
	\label{fig:performance_sim_rate}
\end{figure}

\begin{figure}[!t]
	\centering
	\includegraphics[width=0.8\columnwidth]{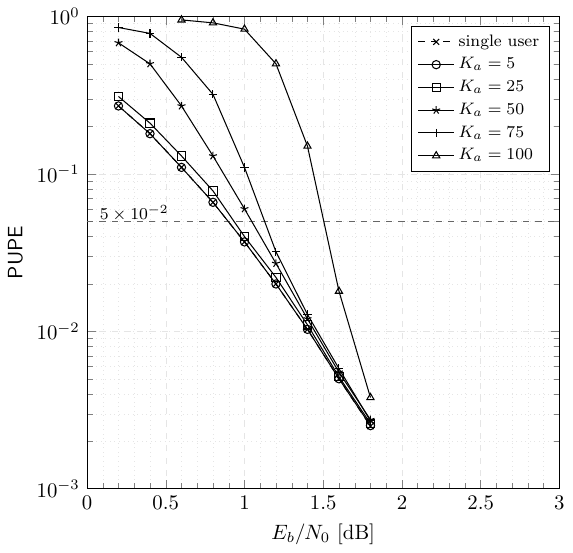}
    \vspace{-2.5mm}
	\caption{\ac{PUPE} vs. \ac{SNR}, for various values of $K_a$. The frame length is $30000$ channel uses.}
	\vspace{-2mm}
	\label{fig:performance_sim_PLR}
\end{figure}

We report Monte Carlo simulation results for a specific \ac{SB-IDMA} configuration, originally designed in \cite{whitepaper24}. The results are obtained by setting the \ac{UMAC} frame length to $30000$ channel uses. The preamble length is $\npre=275$, and the dictionary is composed of $2048$ preambles. The preambles have been generated randomly, with \ac{i.i.d.} entries drawn from a Gaussian distribution with unitary variance. Each message is $100$ bits long, and it is encoded with a $(1000,100)$ CRC-aided (CA) polar code \cite{Ari09,TV15} from the \ac{5GNR} standard. The CA polar code is concatenated with an inner $(4,1)$ repetition code, resulting in an overall blocklength $n_c = 4000$.  Each PO has a dimension of $50$ channel uses. It follows that codewords are partitioned in $N_s = 80$ segments. Preamble detection is performed by \ac{OMP} \cite{TroppOMP}. At the decoder input, the \acp{LLR} associated with the $4$ copies of each codeword bit are combined, and provided to the input of an \ac{SCL} decoder with list size $128$. The performance of this \ac{SB-IDMA} configuration is reported in Fig.~\ref{fig:performance_sim_rate}, in terms of number of active users vs. \ac{SNR}, for a target \ac{PUPE} of $5\times 10^{-2}$. On the same chart, the finite-length achievability bound of \cite{Pol17} is provided. The performance of \ac{SB-IDMA} can be divided into two regions. Up to moderate-low loads (e.g., $K_a \leq 75$), the \ac{SNR} required to achieve the target error probability is almost unaffected by adding more users, and it is close to the one required by a single user transmission w/o multiuser interference. At larger loads ($K_a>75$), the slope of the performance curve visibly changes. Increasing the load beyond this point results in a tangible increase of the required \ac{SNR}. A quick look at Fig.~\ref{fig:performance_sim_PLR} reveals the nature of this phenomenon. The chart reports the \ac{PUPE} as a function of the \ac{SNR}, for different values of $K_a$. At low loads, the \ac{PUPE} converges to the single user performance above the target \ac{PUPE}, or slightly below it. At larger loads, the performance curves display a different behavior: The \ac{PUPE} remains fairly large as the \ac{SNR} grows, then it suddenly drops, almost matching the single user error probability. 
However, this happens at \ac{PUPE} values that are much below the target error probability, hence crossing the target at large \ac{SNR}. This behavior, which is common to various massive \ac{MAC} settings with finite-length messages \cite{KP21,BC02}, will be analyzed in Section \ref{sec:DE}.

\section{Density Evolution Analysis}\label{sec:DE}

The \ac{SB-IDMA} scheme described in Section~\ref{sec:SB-IDMA} can be divided into two phases. The first phase (preamble transmission) is clearly unsourced. Upon detecting the preambles transmitted by the active users, the receiver is aware of the access patterns chosen by the users for data transmission in the second phase. Hence, from a receiver viewpoint, the second phase can be treated as a coordinated \ac{MAC} protocol, where (if all active users pick distinct preambles/access patterns) each user employs a different codebook. In the following, we focus on the analysis of the second phase only, under some simplifying assumptions. First, we assume that all preambles are correctly detected, and no preamble collisions occur. Second, we assume that each preamble points not only to an access pattern but also to a specific block code from a pool of $|\pset|$ codes. Furthermore, access patterns are drawn uniformly at random from the set of all possible $N_s$-tuples of distinct elements in $\{1,2,\ldots,N\}$. Finally, we assume that decoding of a user terminal can result in a correct decoding of the user message, or in a decoding failure that is declared by the decoder via an error flag (no undetected errors). Under these assumptions, we analyze, in the following, the behavior of the \ac{TIN}-SIC receiver in the limit of large frame lengths. To ease analytic treatment, we consider the block codes in the pool to be random. 

\subsection{Graphical Model of the \ac{TIN}-\ac{SIC} Receiver}\label{sec:DE:graphical_model}

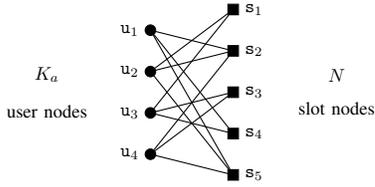
\begin{figure}
	\centering
	\input{graph.tikz}
    \vspace{-1.5mm}
	\caption{Bipartite graph representation of an \ac{SB-IDMA} \ac{UMAC} frame. Circles denote \acp{UN}, and squares denote \acp{SN}.}
	\vspace{-2mm}
\label{fig:bipartite}
\end{figure}

We model the \ac{TIN}-\ac{SIC} decoding process described in Section \ref{sec:SB-IDMA:RX} by means of a bipartite graph \cite{Liv11}. We define a graph $\graph$ with a set of $K_a$ \acp{UN} $\un_1, \un_2, \ldots, \un_{K_a}$, and set of $N$ \acp{SN} $\sn_1, \sn_2, \ldots, \sn_{N}$ (see Fig.~\ref{fig:bipartite}). The \ac{UN} $\un_i$ represents the $i$th active \ac{UT}, and the \ac{SN} $\sn_j$ represents the $j$th slot (PO) in the \ac{UMAC} frame. An edge connects $\un_i$ to $\sn_j$ if and only if a segment of the codeword of the $i$th \ac{UT} is transmitted in the $j$th slot in the frame. The set of neighbors of a \ac{UN} $\un$ (\ac{SN} $\sn$) is $\neigh{\un}$ ($\neigh{\sn}$).  We denote by $d_u$ the degree of \acp{UN} (note that $d_u = N_s$), and by $\bar{d}_s = d_u K_a/N = \mu n_c$ the average degree of the \acp{SN}.

We are interested in the regime where both $N$ and $K_a$ grow linearly in the frame length $n$. This can be achieved by fixing the user density $\mu$, slot size $\npo$, and the blocklength $n_c$ of $\code$. Under these conditions, the number of slots is $N = n/\npo$, and the number of active users is $K_a = \mu n$. The number of segments $N_s = n_c / \npo$ (and, hence, the \ac{UN} degree) is constant.  As $n \to \infty$, \ac{SN} degrees are Poisson-distributed with mean $\bar{d}_s$.

The iterative \ac{TIN}-\ac{SIC} decoding process can be described as a message-passing algorithm over the graph. At the $\ell$th iteration, we denote by $\mzero{\sn_j}{\ell}$ the residual observation at the $j$th slot, by $\msg{\sn_j}{\un_i}{\ell}$ the message sent by $\sn_j$ towards $\un_i$, and by $\msg{\un_i}{\sn_j}{\ell}$ the message sent by $\un_i$ towards $\sn_j$. We introduce the decoding function $\dec{\msgdum{\sn}{\un_i},\sn \in \neigh{\un_i}}$, defining the decoding operation at the generic $i$th \ac{UT} (\ac{UN}). The decoding function can return a codeword or an error flag $\flag$ to denote a decoding failure. If decoding succeeds, the output of the function is a vector $\bm{x}^{(i)} = \left(\segx{x}{1},\segx{x}{2},\ldots,\segx{x}{N}\right)$, where $\segx{x}{j}$ is the interference contribution of the $i$th user transmission in the $j$th slot. The state of the $i$th \ac{UN} is tracked through the binary variable $\state(\un_i) \in \{\missing,\decoded\}$.
In the initialization, we set
$\mzero{\sn_j}{0} = \seg{y}{j}$ for $j=1,2,\ldots,N$ and $\state(\un_i)=\missing$ for $i=1,2,\ldots,K_a$. During the $\ell$th iteration, the \ac{SN}-to-\ac{UN} step 
\begin{equation}
	\msg{\sn_j}{\un_i}{\ell} = \mzero{\sn_j}{\ell-1} \label{eq:snstep}
\end{equation}
is performed for all $\un_i \in \neigh{\sn_j}, j = 1,2,\ldots,N$. The \ac{UN}-to-\ac{SN} step works as follows. At the \ac{UN} $\un_i$, $i=1,2,\ldots,K_a$, if $\state(\un_i)=\decoded$, then 
\begin{equation}
	\msg{\un_i}{\sn_j}{\ell} = \segx{x}{j} \label{eq:unstep1}
\end{equation}
for all $\sn_j\in \neigh{\un_i}$.
Otherwise, decoding via $\mathsf{dec}\big[\msg{\sn}{\un_i}{\ell},\sn \in \neigh{\un_i}\big]$
is attempted at the \ac{UN}. If decoding succeeds, then the output messages are again set as in \eqref{eq:unstep1}, and the \ac{UN} state is updated to $\state(\un_i)=\decoded$. If decoding fails,
\begin{equation}
	\msg{\un_i}{\sn_j}{\ell} = \flag \label{eq:unstep2}
\end{equation}
for all $\sn_j\in \neigh{\un_i}$. Finally, in the interference cancellation step, the residual observation at each \ac{SN} is updated as
\begin{equation}
	\mzero{\sn_j}{\ell} = \seg{y}{j} -\!\!\! \sum_{\substack{\un \in \neigh{\sn_i}\\ \msg{\un}{\sn_j}{\ell}\neq \flag}} \!\!\!\msg{\un}{\sn_j}{\ell}.\label{eq:ICstep}
\end{equation}
The receiver iterates \eqref{eq:snstep} - \eqref{eq:ICstep} for a maximum number of iterations. 
Some observations follow. First, the decoder description could be simplified by casting the iterative \ac{SIC} process in the form of a peeling process, where \acp{UN} (and their attached edges) are removed from the graph whenever a \ac{UN} is decoded. The message passing framework defined by \eqref{eq:snstep} - \eqref{eq:ICstep} is, nevertheless, instrumental to the \ac{DE} analysis of Section \ref{sec:DE:analysis}. Second, the message-passing algorithm described above does not rely on the propagation of extrinsic information. To enable \ac{DE} analysis, in Section \ref{sec:DE:analysis} the algorithm will be modified to comply with extrinsic messages passing.

\subsection{Analysis}\label{sec:DE:analysis}  
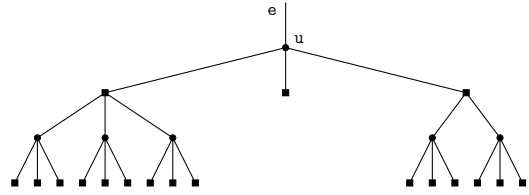
\begin{figure}
	\centering
	\input{tree.tikz}
    \vspace{-1mm}
\caption{Depth-$3$ tree-like neighborhood of a degree-$4$ \ac{UN} $\un$, w/o the edge $\mathtt{e}$.} 
\vspace{-2mm}
\label{fig:tree}
\end{figure}

Consider a generic \ac{UN} $\un$, and pick an arbitrary edge $\mathtt{e}$ connected to $\un$. Let us unfold the depth-$t$ (for some fixed $t$) neighborhood of $\un$, with the exclusion of the edge $\mathtt{e}$. In the limit for $n\to\infty$, the neighborhood is tree-like with high probability. An example of the depth-$3$ neighborhood of a \ac{UN} $\un$, with the exclusion of the edge $\mathtt{e}$, is depicted in Fig.~\ref{fig:tree}. Each \ac{UN} in the tree receives residual observations from its child \acp{SN}, and propagates an estimate of its interference contribution (or an error flag) to the parent \ac{SN}. We assume the following modification of the algorithm of Section~\ref{sec:DE:graphical_model}. Each \ac{UN} in the tree attempts the decoding by using only the residual observations of the children \acp{SN}, i.e., it disregards the residual observation at the parent \ac{SN}. That is, the message from a \ac{UN} $\un_i$ to a parent node $\sn_j$ is the result of the decoding operation 
\begin{equation}
\dec{\msgdum{\sn}{\un_i},\sn \in \neigh{\un_i}\setminus\{\sn_j\}}.\label{eq:extdec}
\end{equation}
This modification allows us to perform the \ac{DE} analysis by forcing the \ac{UN} to output extrinsic information \cite{JPH17}. A drawback of this choice is that the local \ac{UN} decoder will perform sub-optimally. In fact,  \eqref{eq:extdec} is equivalent to decoding a punctured version of the local code. Nevertheless, if the \ac{UN} degree is sufficiently large, the fraction of punctured bits  ($1/d_u$) is negligible (in the example of Section~\ref{sec:SB-IDMA:PUPE}, $d_u = 80$ and, hence, puncturing amounts to $1.25 \%$ of the bits).

We are interested in tracking, over the tree, the average probability of decoding failure $\epsilon$ at the \ac{UN} output. Consider a generic \ac{SN} $\sn$ in the tree, and denote the number of its children \acp{UN} by the \ac{r.v.} $D\sim\poisson{\bar{d}_s}$. Since some of the children \acp{UN} may have been decoded successfully with probability $1-\epsilon_{\ell-1}$, the number of interferers ``seen'' by the parent \ac{UN} $\un$, over the slot associated to $\sn$, can be modeled by a \ac{r.v.} $G\sim\poisson{\bar{d}_s\epsilon_{\ell}}$. Consider now a generic parent \ac{UN} $\un$. Denote by $\bm{G} = (G_1, G_2, \ldots, G_{d_u-1})$ a vector containing the number of residual interferers ``seen'' by $\un$ in each of the slots corresponding to its $d_u-1$ children \acp{SN}. For a given \ac{SNR}, and for a residual interference vector $\bm{G}$, we denote the average decoding error probability according to the extrinsic rule of \eqref{eq:extdec} by $\varphi(\bm{G};E_b/N_0))$. By averaging over $\bm{G}$, the decoding failure probability is
\begin{align}
	\epsilon_\ell &= \expect{\varphi(\bm{G};E_b/N_0)} \label{eq:DE1}\\
	& =f(\epsilon_{\ell-1};\bar{d}_s,E_b/N_0) \label{eq:DE2}
\end{align}
where $G_1, G_2, \ldots, G_{d_u-1}$ are \ac{i.i.d.} Poisson \acp{r.v.} with mean $\bar{d}_s\epsilon_{\ell-1}$. 
For the evaluation of the function $f(\cdot)$, we resort to random coding and to the \ac{DT} bound of \cite{PPV10} (see Appendix~\ref{app:decoding}). The analysis proceeds by studying the fixed points of the recursion \eqref{eq:DE2} initalized with $\epsilon_0 = 1$.

\section{Numerical Example and Concluding Remarks}\label{sec:example}
\begin{figure}[t]
	\centering
	\includegraphics[height=0.5\columnwidth]{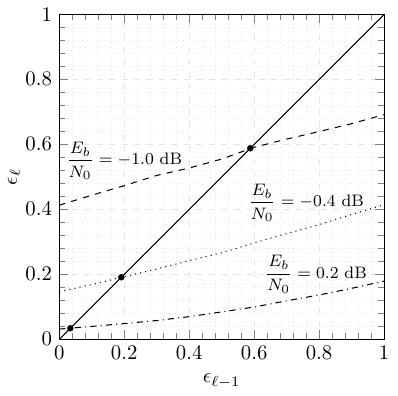}\includegraphics[height=0.5\columnwidth]{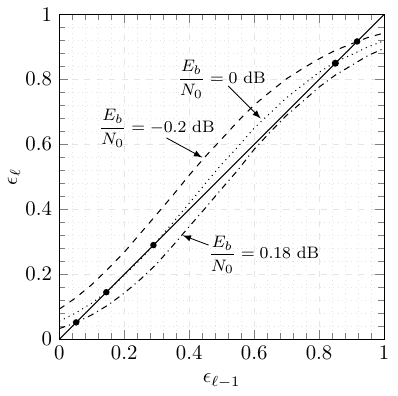}
    \vspace{-2.5mm}
	\caption{Fixed points of \eqref{eq:DE2}, $\mu = 8.41\times 10^{-4}$ (left) and $\mu = 3.40\times 10^{-3}$ (right).}
	\vspace{-1.5mm}
	\label{fig:K25100}
\end{figure}

\begin{figure}
	\centering
	\includegraphics[width=0.8\columnwidth]{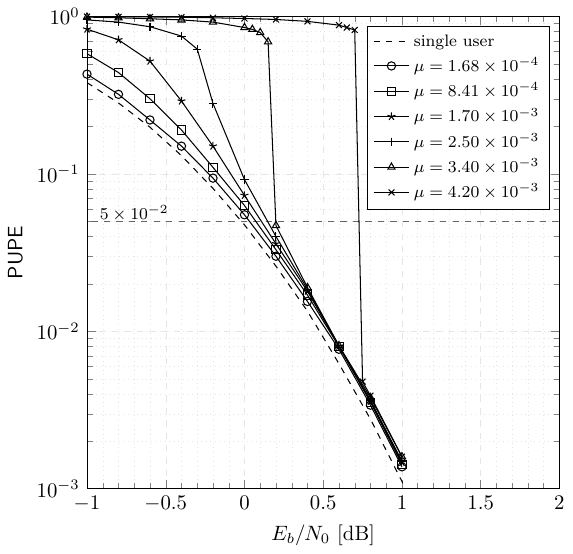}
    \vspace{-2.5mm}
	\caption{\ac{PUPE} prediction according to the \ac{DE} analysis.}
	\vspace{-3 mm}
	\label{fig:DE}
\end{figure}

We consider a setting similar to the one described in Section~\ref{sec:SB-IDMA:PUPE}. In particular, each active user encodes an information message of $k=100$ bits by means of a random code with blocklength $4000$ channel uses. The codebooks are Gaussian. Each codeword is partitioned in $N_s=80$ segments of $\npo=50$ channel uses each. Fig.~\ref{fig:K25100} (left) reports---in an EXIT chart fashion---the function \eqref{eq:DE2} for a user density $\mu = 8.41 \times 10^{-4}$. Such user density, in the finite-length setting of Section~\ref{sec:SB-IDMA:PUPE}, would result in $K_a = 25$. The plot depicts the function at three different values of \ac{SNR}. In all cases, there is a unique fixed point, yielding an error probability that is fairly close to the single user one (given by the intercept, at $\epsilon_{\ell-1}=0$, of the curves with the $y$-axis). A qualitative different behavior can be observed at a much larger user density. In Fig.~\ref{fig:K25100} (right), the recursion \eqref{eq:DE2} is depicted, for $\mu = 3.40 \times 10^{-3}$. Such user density, in the finite-length setting of Section~\ref{sec:SB-IDMA:PUPE}, would result in $K_a =100$. Again, the plot is provided for three different \acp{SNR}. At the two lowest \acp{SNR}, additional fixed points emerge, in the high error probability regime. Here, the iterative \ac{TIN}-\ac{SIC} receiver gets stuck at an error probability that is much larger than the one of a single-user decoder. At the highest \ac{SNR} ($0.18$ dB), the fixed points at high error probability disappear, leaving place to a unique fixed point, yielding an error probability that is tight on the one of a single-user decoder. This suggests that, between the intermediate \ac{SNR} ($0$ dB) and the largest \ac{SNR} ($0.18$ dB), a phase transition occurs, which allows the iterative \ac{SIC} process to reach a performance that is close to the single user decoder one. This phenomenon is clearly visible in Fig.~\ref{fig:DE}, where the error probability achieved at the first fixed point for various user densities is depicted as a function of the \ac{SNR}. Following the observation of Fig.\ref{fig:K25100}, we see that, at low user density, the \ac{DE} analysis predicts a \ac{PUPE} performance that converges to the single user error probability. At larger user densities, we observed a high error probability at low \ac{SNR}, which is caused by the emergence of a high error probability fixed point in the recursion \eqref{eq:DE2}. As the \ac{SNR} grows larger, the fixed point disappears, and the performance curve ``jumps'' to much lower values of the \ac{PUPE}, almost matching the the single user error probability. A similar phenomenon was observed in the context of iterative \ac{MUD} schemes in \cite{BC02}. 

Note that, despite all the assumptions made in the \ac{DE} analysis of Section~\ref{sec:DE}, the analysis can still be used as a proxy for the performance of the actual scheme. In fact, by comparing the result depicted in Fig.\ref{fig:DE} with the simulation results of Fig.~\ref{fig:performance_sim_PLR}, and by accounting for the energy overhead introduced by the preambles in the simulation setting (which amounts to approx. $0.3$ dB), we can observe that the numerical results of Fig.~\ref{fig:performance_sim_PLR} are still fairly close to the \ac{DE} analysis of Fig.\ref{fig:DE}.

\section*{Acknowledgment}
The authors thank Yury Polyanskiy for the insightful discussions that contributed to the analysis presented in this paper.

\appendix \label{app:decoding}

	Consider a random Gaussian codebook $\code$ with $M=2^k$ codewords. Codeword symbols are distributed as $\mathcal{N}(0,1)$. Upon transmission of a random codeword $\bm{X}$, the decoder observes $\bm{Y}$. Let us partition $\bm{X}$ into $d_u$ length-$\npo$ segments $\bm{X} = (\seg{X}{1},\seg{X}{2},\ldots,\seg{X}{d_u})$. Similarly, let $\bm{Y} = (\seg{Y}{1},\seg{Y}{2},\ldots,\seg{Y}{d_u})$ where
	$\seg{Y}{i} = \seg{X}{i} + \seg{Z}{i} + \seg{V}{i}$ 
	for $i=1,2,\ldots,d_u$, with $\seg{Z}{i}$ composed by $\npo$ \ac{i.i.d.} \ac{AWGN} terms with zero-mean and variance $\sigma^2$, and where $\seg{V}{i}$ is composed by $\npo$ \ac{i.i.d.} Gaussian interference terms with zero-mean and variance $G_i$. Consider the observation of the last segment to be erased, at the decoder input. The decoder has therefore access only to $\bm{Y}_{\sim d_u} = (\seg{Y}{1},\seg{Y}{2},\ldots,\seg{Y}{d_u-1})$. The information density associated with $\bm{X}$ and $\bm{Y}_{\sim d_u}$ is
	\begin{equation}
	\iota(\bm{X},\bm{Y}_{\sim d_u}) = \log_2 \!\frac{p(\bm{Y}_{\sim d_u}|\bm{X})}{p(\bm{Y}_{\sim d_u})} = \sum_{i=1}^{d_u-1}\! \log_2 \!\frac{p(\bm{Y}_i|\bm{X}_i)}{p(\bm{Y}_i)}
	\end{equation}
	where, in the rightmost summation, the term for $i=d_u$ is absent due to the missing observation $\seg{Y}{d_u}$ (erasure). For given $\bm{G}=(G_1,G_2, \ldots, G_{d_u-1})$, the random code average block error probability satisfies \cite[Theorem 17]{PPV10}
	\begin{equation}
	\varphi(\bm{G}) \leq \expect{2^{-\left[\iota(\bm{X},\bm{Y}_{\sim d_u}) - \log_2\frac{M-1}{2}\right]^+}\Big| \bm{G}}. \label{eq:DT}
	\end{equation}
	An estimate of \eqref{eq:DE2} under random coding with Gaussian codebooks can be obtained by averaging \eqref{eq:DT} over the distribution of $G_1,G_2, \ldots, G_{d_u-1}$ as in \eqref{eq:DE1}.

\end{document}

%% file: diagramm.tikz.tex
\begin{tikzpicture}[scale = 0.65, every node/.style={scale = 0.65}]
	\draw[-] (0,0) -- (2,0);
	\draw[-] (2,0) -- (10,0);
	\foreach \i in {0,2,10} {
		\draw (\i,-0.2) -- (\i,0.2);
	}
	
	%\fill[black!5] (0.05,1.5) rectangle (1.95,2.5);
	%\fill[black!5] (2.05,1.5) rectangle (9.95,2.5);
	%\fill[black!5] (0.05,-0.7) rectangle (1.95,-1.3);
	%\fill[black!5] (2.05,-0.7) rectangle (9.95,-1.3);
	
	\node[align=center] at (1,-0.7) {PRACH};
	\node[align=center] at (1,1.7) {preamble\\ $\npre$ cu};
	\node[align=center] at (6,-0.7) {$N$ POs};
	\node[align=center] at (6,1.7) {data part\\ $N\times\npo$ cu};
	
	\foreach \i in {3,...,9} {
		\draw (\i,-0.1) -- (\i,0.1);
	}
	\foreach \i in {1,...,2} {
		\node[align=center,scale = 0.8] at (\i+1.5,-0.3) {PO $\i$};
	}
	\node[align=center] at (6,-0.3) {$\dots$};
	\node[align=center,scale = 0.8] at (9.5,-0.3) {PO $N$};
	
	\draw[pattern=north east lines] (0,0.5) rectangle (2,0.8);
	\draw[pattern=north east lines] (3,0.5) rectangle (4,0.8);
	\draw[pattern=north east lines] (5,0.5) rectangle (6,0.8);
	\draw[pattern=north east lines] (6,0.5) rectangle (7,0.8);
	\draw[pattern=north east lines] (9,0.5) rectangle (10,0.8);
	
\end{tikzpicture}

%% file: graph.tikz.tex
\begin{tikzpicture}[scale=0.55,every node/.style={scale=0.7}]
	\vnode{0,3}{v1}
	\vnode{0,2}{v2}
	\vnode{0,1}{v3}
	\vnode{0,0}{v4}
	
	\node at (-0.5,3) {$\un_1$};
	\node at (-0.5,2) {$\un_2$};
	\node at (-0.5,1) {$\un_3$};
	\node at (-0.5,0) {$\un_4$};
	
	\cnode{2,3.5}{c1}
	\cnode{2,2.5}{c2}
	\cnode{2,1.5}{c3}
	\cnode{2,0.5}{c4}
	\cnode{2,-0.5}{c5}
	
	\node at (2.5,3.5) {$\sn_1$};
	\node at (2.5,2.5) {$\sn_2$};
	\node at (2.5,1.5) {$\sn_3$};
	\node at (2.5,0.5) {$\sn_4$};
	\node at (2.5,-0.5) {$\sn_5$};
	
	\draw (v1) -- (c2);
	\draw (v1) -- (c4);
	\draw (v1) -- (c5);
	
	\draw (v2) -- (c1);
	\draw (v2) -- (c2);
	\draw (v2) -- (c5);
	
	\draw (v3) -- (c1);
	\draw (v3) -- (c3);
	\draw (v3) -- (c4);
	
	\draw (v4) -- (c2);
	\draw (v4) -- (c3);
	\draw (v4) -- (c5);
	
	\node[align=center] at (-2.5,1.5) {$K_a$\\[3mm] user nodes};
	\node[align=center] at (4.5,1.5) {$N$\\[2mm] slot nodes};
\end{tikzpicture}

%% file: tree.tikz.tex
\begin{tikzpicture}[scale=0.6,every node/.style={scale=0.4}]	
	%\draw[draw=white,fill = black!10] (-4.2,-1.2) rectangle (4.2,0.2);	
	%\draw[draw=white,fill = black!10] (-6.2,-3.2) rectangle (-4.8,-1.8);	
	%\draw[draw=white,fill = black!10] (-4.7,-3.2) rectangle (-3.3,-1.8);	
	%\draw[draw=white,fill = black!10] (-3.2,-3.2) rectangle (-1.8,-1.8);
	%\draw[draw=white,fill = black!10] (-4.7,-3.2) rectangle (-3.3,-1.8);	
	%\draw[draw=white,fill = black!10] (2.55,-3.2) rectangle (3.95,-1.8);		
	%\draw[draw=white,fill = black!10] (4.05,-3.2) rectangle (5.45,-1.8);	
	
	\vnode{0,0}{0v};
	\draw[thin, solid, black] (0v) -- ++(0, +1);\node at (0.3,0.2) {\LARGE $\un$};\node at (-0.3,0.8) {\LARGE $\mathtt{e}$};
	\cnode{-4,-1}{0c1};
	\cnode{0,-1}{0c2};
	\cnode{4,-1}{0c3};		
	\draw[thin] (0v) -- (0c1);
	\draw[thin] (0v) -- (0c2);
	\draw[thin] (0v) -- (0c3);

	\vnode{-5.5,-2}{0c1v1};	
	\vnode{-4,-2}{0c1v2};	
	\vnode{-2.5,-2}{0c1v3};
	\draw[thin] (0c1) -- (0c1v1);
	\draw[thin] (0c1) -- (0c1v2);
	\draw[thin] (0c1) -- (0c1v3);
	
	\vnode{3.25,-2}{0c3v1};
	\vnode{4.75,-2}{0c3v2};
	\draw[thin] (0c3) -- (0c3v1);
	\draw[thin] (0c3) -- (0c3v2);
	
	\cnode{-6.0,-3}{0c1v1c1};
	\cnode{-5.5,-3}{0c1v1c2};	
	\cnode{-5.0,-3}{0c1v1c3};
	\draw[thin] (0c1v1) -- (0c1v1c1);
	\draw[thin] (0c1v1) -- (0c1v1c2);
	\draw[thin] (0c1v1) -- (0c1v1c3);	
	
	\cnode{-4.5,-3}{0c1v2c1};	
	\cnode{-4.0,-3}{0c1v2c2};	
	\cnode{-3.5,-3}{0c1v2c3};
	\draw[thin] (0c1v2) -- (0c1v2c1);
	\draw[thin] (0c1v2) -- (0c1v2c2);
	\draw[thin] (0c1v2) -- (0c1v2c3);		
	
	\cnode{-3.0,-3}{0c1v3c1};	
	\cnode{-2.5,-3}{0c1v3c2};	
	\cnode{-2.0,-3}{0c1v3c3};
	\draw[thin] (0c1v3) -- (0c1v3c1);
	\draw[thin] (0c1v3) -- (0c1v3c2);
	\draw[thin] (0c1v3) -- (0c1v3c3);		
	
	\cnode{2.75,-3}{0c3v1c1};	
	\cnode{3.25,-3}{0c3v1c2};	
	\cnode{3.75,-3}{0c3v1c3};
	\draw[thin] (0c3v1) -- (0c3v1c1);
	\draw[thin] (0c3v1) -- (0c3v1c2);
	\draw[thin] (0c3v1) -- (0c3v1c3);		
	
	\cnode{4.25,-3}{0c3v2c1};	
	\cnode{4.75,-3}{0c3v2c2};	
	\cnode{5.25,-3}{0c3v2c3};
	\draw[thin] (0c3v2) -- (0c3v2c1);
	\draw[thin] (0c3v2) -- (0c3v2c2);
	\draw[thin] (0c3v2) -- (0c3v2c3);

\end{tikzpicture}

%% file: SPAWC25_DE_SB_IDMA.bbl
\vfill	

\begin{thebibliography}{10}
	\providecommand{\url}[1]{#1}
	\csname url@samestyle\endcsname
	\providecommand{\newblock}{\relax}
	\providecommand{\bibinfo}[2]{#2}
	\providecommand{\BIBentrySTDinterwordspacing}{\spaceskip=0pt\relax}
	\providecommand{\BIBentryALTinterwordstretchfactor}{4}
	\providecommand{\BIBentryALTinterwordspacing}{\spaceskip=\fontdimen2\font plus
		\BIBentryALTinterwordstretchfactor\fontdimen3\font minus
		\fontdimen4\font\relax}
	\providecommand{\BIBforeignlanguage}[2]{{%
			\expandafter\ifx\csname l@#1\endcsname\relax
			\typeout{** WARNING: IEEEtran.bst: No hyphenation pattern has been}%
			\typeout{** loaded for the language `#1'. Using the pattern for}%
			\typeout{** the default language instead.}%
			\else
			\language=\csname l@#1\endcsname
			\fi
			#2}}
	\providecommand{\BIBdecl}{\relax}
	\BIBdecl
	
	\bibitem{5GNR16}
	\emph{{5GNR: Medium Access Control (MAC) protocol specification}}, 3GPP Std. TS
	138.321, Rev. 16.1.0, Jul. 2020.
	
	\bibitem{LP24}
	G.~Liva and Y.~Polyanskiy, ``{Unsourced Multiple Access: A Coding Paradigm for
		Massive Random Access},'' \emph{Proc. {IEEE}}, vol. 112, no.~9, pp.
	1214--2256, Sep. 2024.
	
	\bibitem{whitepaper24}
	\BIBentryALTinterwordspacing
	P.~{Agostini \textit{et al.}}, ``Evolution of the {5GNR} two-step random access
	towards {6G} unsourced {MAC},'' Apr. 2024, {Report V.1.0}. [Online].
	Available: \url{https://arxiv.org/abs/2405.03348}
	\BIBentrySTDinterwordspacing
	
	\bibitem{Pol17}
	Y.~Polyanskiy, ``A perspective on massive random-access,'' in \emph{Proc.
		{IEEE} Int. Symp. Inf. Theory}, {Aachen, Germany}, 2017.
	
	\bibitem{Rob75}
	L.~G. Roberts, ``{ALOHA packet system with and without slots and capture},''
	\emph{ACM SIGCOMM Comput. Commun. Rev.}, vol.~5, no.~2, p. 28–42, Apr.
	1975.
	
	\bibitem{PAV+22}
	A.~K. Pradhan, V.~K. Amalladinne, A.~Vem, K.~R. Narayanan, and J.-F.
	Chamberland, ``{Sparse IDMA: A Joint Graph-Based Coding Scheme for Unsourced
		Random Access},'' \emph{{IEEE} Trans. Commun.}, vol.~70, no.~11, pp.
	7124--7133, Nov. 2022.
	
	\bibitem{RU08}
	T.~Richardson and R.~Urbanke, \emph{Modern coding theory}.\hskip 1em plus 0.5em
	minus 0.4em\relax Cambridge University Press, 2008.
	
	\bibitem{Ari09}
	E.~Arikan, ``Channel polarization: A method for constructing capacity-achieving
	codes for symmetric binary-input memoryless channels,'' \emph{{IEEE} Trans.
		Inf. Theory}, vol.~55, no.~7, pp. 3051--3073, Jul. 2009.
	
	\bibitem{TV15}
	I.~Tal and A.~Vardy, ``List decoding of polar codes,'' \emph{{IEEE} Trans. Inf.
		Theory}, vol.~61, no.~5, pp. 2213--2226, May 2015.
	
	\bibitem{TroppOMP}
	J.~A. Tropp and A.~C. Gilbert, ``Signal recovery from random measurements via
	orthogonal matching pursuit,'' \emph{{IEEE} Trans. Inf. Theory}, vol.~53,
	no.~12, pp. 4655--4666, Dec. 2007.
	
	\bibitem{KP21}
	S.~S. Kowshik and Y.~Polyanskiy, ``Fundamental limits of many-user {MAC} with
	finite payloads and fading,'' \emph{{IEEE} Trans. Inf. Theory}, vol.~67,
	no.~9, pp. 5853--5884, Sep. 2021.
	
	\bibitem{BC02}
	J.~Boutros and G.~Caire, ``Iterative multiuser joint decoding: Unified
	framework and asymptotic analysis,'' \emph{{IEEE} Trans. Inf. Theory},
	vol.~48, no.~7, pp. 1772--1793, Jul. 2002.
	
	\bibitem{Liv11}
	G.~Liva, ``Graph-based analysis and optimization of contention resolution
	diversity slotted {ALOHA},'' \emph{{IEEE} Trans. Commun.}, vol.~59, no.~2,
	pp. 477--487, Feb. 2011.
	
	\bibitem{JPH17}
	Y.-Y. Jian, H.~D. Pfister, and K.~R. Narayanan, ``Approaching capacity at high
	rates with iterative hard-decision decoding,'' \emph{{IEEE} Trans. Inf.
		Theory}, vol.~63, no.~9, pp. 5752--5773, Sep. 2017.
	
	\bibitem{PPV10}
	Y.~Polyanskiy, H.~V. Poor, and S.~Verd{\'u}, ``Channel coding rate in the
	finite blocklength regime,'' \emph{{IEEE} Trans. Inf. Theory}, vol.~56,
	no.~5, pp. 2307--2359, May 2010.
	
\end{thebibliography}
